      [1999/12/01 v1.4c Il Nuovo Cimento]
\documentclass{cimento}

\title{From Vector Meson Dominance to Quark-Hadron Duality}
\author{A.~Bramon\from{ins:x}}
\instlist{\inst{ins:x} Dept. de F{\'{\i}}sica, Universitat Aut\`onoma de Barcelona, \\
08193,~Bellaterra,~Barcelona.}
\begin{document}

\maketitle

\begin{abstract}
A short review of the many contributions to hadron physics made by Mario Greco  is presented. The review roughly covers his production between 1971 and 1974, just before the advent of QCD, when quark-model ideas, duality principles and vector meson dominance were widely accepted, developed and applied. The present author had the privilege to collaborate with Mario in most of these contributions and looking backward in time to remember the good old days we spent together in Frascati has been a great pleasure.  
\end{abstract}

\section{Vector Mesons and Vector Meson Dominance}

In 1960 J.J. Sakurai published an influential paper~\cite{ref:sak},  \TITLE{Theory of Strong Interactions}, proposing the still to be discovered vector mesons ($J^{PC}=1^{--}$) as gauge bosons of the strong interactions of hadrons. In this way, the successful gauge principle of QED, generating universal couplings via covariant derivatives, was exported to the much more involved field of hadron physics. The special role given to vector mesons was in clear contrast with the `hadronic democracy' ideas widely accepted at that time and during the forthcoming 1960's. Indeed, all hadrons were expected to be treated at the same level using the unitarity and analyticity properties of the S-matrix, organizing hadron resonances along Regge trajectories and unitary symmetry multiplets of SU(3), or proposing crossing-symmetric amplitudes, as in the Veneziano model.  

Low-mass vector mesons had been predicted just a few years before Sakurai's paper in order to understand the structure and size of protons and neutrons~\cite{ref:predict}. The pionic cloud surrounding the nucleons was assumed to be responsible for their structure, usually expressed in terms of the so called electromagnetic form factors. Form factor data required, at least, two low-mass vector mesons: an isoscalar, $I=0$, $\omega$-resonance and 
the neutral member, $\rho^0$, of an isovector triplet, $I=1$.  Both states, $\omega$~\cite{ref:omega} and $\rho$~\cite{ref:rho},  were experimentally identified in 1961 short after Sakurai's proposal~\cite{ref:sak}. Other low-mass vector resonances were discovered and organized in a complete SU(3)-nonet which became  paradigmatic in applying SU(3) arguments. In particular, the almost ideal mixing between the  $\omega$ and $\phi$ isoscalar states and the so-called Zweig-rule,  strongly suggested the existence of quarks in 1964 and, almost immediately, the `quark model' became an extremely popular and successful tool in hadron physics.

It was soon realized that Sakurai's ideas could explain only a few features of the strictly {\it strong} interactions of hadrons and that a new theory 
--QCD, to be formulated some ten years later-- 
was needed. However, restricting to the more specific context of the {\it electromagnetic} interactions of hadrons, the gauge principles of ref.~\cite{ref:sak} turned out to be extremely successful. The key ingredient, proposed in 1961 by Gell-Mann and Zachariasen~\cite{ref:VMD}, was the complete vector meson dominance (VMD) of the electromagnetic form factors of hadrons. On the one hand, vectors mesons were universally coupled to the various hadrons --thus making the model extremely predictive-- and, on the other, the three relevant vector mesons, 
$V=\rho^0$, $\omega$ and $\phi$, were coupled to the photon via the constants
\begin{eqnarray}
\label{emf}
{e m^2_V\over f_V}\, \;\; \;\; \rm with \;\;V=\rho^0 ,\; \omega ,\; \phi .
\end{eqnarray}
The values of the masses, $m^2_V$, and coupling constants, $f_V$, were soon accurately measured by the Orsay, Novosibirsk and Frascati 
$e^+ e^-$ storage rings and found to be in nice agreement with quark model arguments. 

One of the first and cleanest applications of VMD was the establishment of the following relation between the total photoproduction cross section off nucleons, $\sigma_{\rm tot}(\gamma p) $, and the forward photoproduction cross section of $V$'s, ${{\rm d}\sigma_0 \over {\rm dt}} (\gamma p \to V p)$, 
\begin{eqnarray}
\label{gp}
\sigma_{\rm tot}(\gamma p) = \sum_{V= \rho^0, \omega, \phi}  {4\pi e \over f_V} 
\sqrt{{1 \over 1+\eta^2_V }{{\rm d}\sigma_0 \over {\rm dt}} (\gamma p \to V p)},
\end{eqnarray}
which is an immediate consequence of the couplings (\ref{emf}) and the optical theorem (in this context, $\eta_V$ accounts for the real part of the $V$-production amplitude). 
A second, well-known VMD result is illustrated by the following three decay chains 
\begin{eqnarray}
\label{VP}
\omega \to \rho \pi \to \pi^+\pi^-\pi^0 \, , \;\;\; 
\omega \to \pi^0 \rho^0 \to \pi^0 \gamma \, , \;\;\; 
\pi^0 \to \omega \rho^0 \to \gamma \gamma
\end{eqnarray}
which allowed to relate the $\Gamma (\omega \to \pi^+\pi^-\pi^0 )$, $\Gamma (\omega \to \pi^0 \gamma )$ and 
$\Gamma (\pi^0 \to \gamma \gamma)$ decay widths. In all these cases, VMD predictions turned out to be in reasonable agreement with the available data. 
\section{Extended Vector Meson Dominance}

Just after 1970, when more accurate data on the above processes became available, some discrepancies with the VMD predictions started to appear and to be discussed.  Data on the $\sigma_{\rm tot}(\gamma p) $ cross section were found to be some 20\% larger than predicted by the right hand side of eq.~(\ref{gp}). This discrepancy strongly suggested the convenience to extend the sum in this equation to include  further contributions from new, higher-mass vector mesons. This generalized VMD model was proposed by Sakurai and Schildknecht~\cite{ref:ss1} and has recently been reviewed in detail in~\cite{ref:sch}. 

Somewhat earlier and along the same lines, Mario Greco and the present author~\cite{ref:bg1} discussed the convenience to extend VMD with a second SU(3)-nonet of vector mesons, $V^\prime$,  to account for new data on $V \to PPP$, $V \to P\gamma$ and $P \to \gamma \gamma$ decays, where $V$ and $P$ stand for the various members of the vector- and pseudoscalar-meson nonets thus generalizing the two-step processes quoted in (\ref{VP}). A fit to these data led to an estimate of the relevant coupling strengths of the higher-mass vector mesons, $V^\prime$,  and to a few predictions such as
\begin{eqnarray}
\label{VP1}
& &\sigma_{e^+ e^- \to \omega \pi^0 \to \pi^0 \pi^0 \gamma}(m^2_{\rho^\prime}) \simeq {\rm 3\; nb} \\
\label{VP2}
& &\sigma_{e^+ e^- \to \rho \eta \to \pi^+ \pi^- \eta}(m^2_{\rho^\prime}) \simeq {\rm 2\; nb}, 
\end{eqnarray}  
where we assumed a $\rho^\prime$ mass around 1.5 GeV. Such a value was in the mass region being explored those days by the ADONE  $e^+ e^-$ storage ring in Frascati, where Greco's group was placed. The hope was  that our next-door experimental colleagues could confirm these predictions but, due to the smallness of the estimated cross sections, we had to wait for more than 30 years. Only the quite recent measurements from DM2, CMD-2 and SND, which can reasonably be averaged to the peak-values $\sigma_{e^+ e^- \to \omega \pi^0 \to \pi^0 \pi^0 \gamma}(m^2_{\rho^\prime}) \simeq {\rm 1.5\; nb}$ 
and $\sigma_{e^+ e^- \to \rho \eta \to \pi^+ \pi^- \eta}(m^2_{\rho^\prime}) \simeq {\rm 3\; nb}$, have shown some agreement with our rough estimates~(\ref{VP1}) and~(\ref{VP2}). 

Nowadays it's rather well established that the dominant decay mode of the $\rho^\prime$, with a mass around 1.5 GeV, is into four pions and not into the above two channels (\ref{VP1}) and~(\ref{VP2}) to which our extended VMD approach could be directly applied. 
The $e^+ e^- \to \rho^\prime \to \pi^+ \pi^- \pi^+ \pi^-$  was experimentally observed in Frascati in 1971~\cite{ref:bar} and 
further theoretical analyses of this process were discussed in~\cite{ref:bg2} and~\cite{ref:bg3}. Interestingly enough, in~\cite{ref:bg2} M.~Greco insisted in making the final remark that {\it ``it is very tempting to speculate on what can be the overall contribution of [an infinite set of] vector mesons coupled to the photon''.} A naive extrapolation of our results, modifying eq.~(\ref{gp}) through the introduction of an infinity of higher mass vector mesons, led to a surprising good agreement with the available photoproduction data and, more importantly, opened the door to an interesting new idea.   

\section{Vector Meson Dominance, scale invariance and a `new' duality}

Starting in 1971, deep inelastic scattering experiments established the `scale invariant' behaviour of the structure functions of the nucleons. A related behaviour, implying that the total cross section of hadron production in $e^+ e^-$ annihilations, $\sigma_{\rm had}(s)$, scales as $1/s$ for large values of the CM energy $\sqrt{s}$, had been suggested by Bjorken. As a result,  the `quark-parton'  model, and improved versions of it,  were proposed to explain these deep inelastic phenomena in terms of point-like constituents or `partons'. An interesting possibility was to attempt 
{\it a VMD approach to scale invariance.} With this explicit tittle we published a short paper in 1972~\cite{ref:beg3}, which was  further developed and reviewed by Mario somewhat later~\cite{ref:mg1}. The central points of this proposal were favourably accepted and defended, among others, by Sakurai in a nice paper~\cite{ref:sak2}, where our original contribution~\cite{ref:beg3} was reanalyzed and rephrased in terms a `new kind of duality'.  

According to Sakurai's presentation, the assumption in~\cite{ref:beg3} and~~\cite{ref:mg1}  
that the total cross section of hadron production in $e^+ e^-$ annihilation is completely dominated by the formation of vector mesons can be written as  
\begin{eqnarray}
\sigma_{\rm had}(s)
\label{had} 
&=& {12 \pi \over s} \sum_V {m^2_V \Gamma_V \Gamma_{V \to e^+ e^-} \over 
(s-m^2_V)^2 + m^2_V \Gamma^2_V} \\ \nonumber
&=& \sigma_{\mu \, {\rm pair}}(s) \sum_V {3 \over (f^2_V /4\pi)} {m^3_V \Gamma_V  \over 
(s-m^2_V)^2 + m^2_V \Gamma^2_V},
\end{eqnarray}
where we have introduced the $V-\gamma$ coupling (\ref{emf}) and the $\mu^+ \mu^-$-pair cross section at large CM energies, $s>>m^2_\mu$, 
\begin{eqnarray}
 \sigma_{\mu \, {\rm pair}}(s) = 4\pi \alpha^2 / 3s.
\end{eqnarray}
One immediately sees that if  $\sigma_{\rm had}(s) $  at high energies has to behave like $1/s$, the sum in eq.~(\ref{had}) has to be extended to  an infinite series of vector mesons, as already suggested in~\cite{ref:bg2}. Moreover, a very specific relation has to exist between the density of vector meson states 
per unit squared mass interval, $P_V(m^2)= 1/ \Delta m_V^2$, their masses $m_V$ and couplings to the photon $f_V$; namely, 
\begin{eqnarray}
P_V(m^2)m^2_V /f^2_V = m^2_V / \Delta m_V^2 f^2_V = {\rm constant.}
\end{eqnarray}
The other assumptions of our model~\cite{ref:beg3} and~\cite{ref:mg1}
were that (a) the $1/s$ behaviour is obtained on the average even for the prominent low-mass vector mesons (precocious scaling), 
(b) the isovector meson spectrum is given by 
\begin{eqnarray}
m^2_n = m^2_\rho (1+2n) 
\,, \;\; n= 0,\, 1, \, 2, \dots \, ,
\end{eqnarray}
as in the Veneziano model, and (c) the isoscalar sector contributes, as usual, an additional 1/3 to the isovector contribution. 
Taking all this into account, the model proposed in~\cite{ref:beg3} and~~\cite{ref:mg1} reached two relevant predictions: on the one hand, a reasonable description of the nucleon structure functions was achieved and, on the other, the remarkably simple  prediction
\begin{eqnarray}
R \equiv \displaystyle\lim_{s \to \infty} \sigma_{\rm had}(s)/ \sigma_{\mu \, {\rm pair}}(s) = 2\pi / (f^2_\rho /4 \pi) \simeq 2.5,
\end{eqnarray}
which was expected to be valid for large $s$ (but below the opening of the yet undiscovered new flavour's channels) 
and turned out to be quite important for our present purposes, was obtained. 

Indeed, it's this latter equation what probably suggested Sakurai to rephrase our findings as a `new' duality in  
$e^+ e^-$-annihilations into hadrons.  In our model, the numerical prediction $R\simeq 2.5$, which is in good agreement with the data for 
$\sqrt{s}$ between 2 and 3 GeV, follows exclusively from the value of $f_\rho$ (plus analogous isoscalar contributions) which is a low-mass resonance parameter. For asymptotic, large values of $s$, the ratio $R$ and the $1/s$ scaling of the $e^+ e^-$ cross section into hadrons are well predicted by the quark-parton model, much in the same way as in high energy hadron-hadron collisions the relevant asymptotic amplitudes are well described by the exchange of Regge trajectories. The `old' strong interaction duality between the high-energy Regge amplitudes 
(in the $t$ and $u$ channels) and the low-energy resonance formation (in the $s$ channel) thus admits a vivid analogue in $e^+ e^-$ interactions: on the average, the contributions from low-energy vector mesons are dual to the contributions from the corresponding point-like quarks. These contributions should not be added (this would imply a `double counting' of the contributions), they rather satisfy `finite energy sum rules' (see the final paragraphs for some examples) as in the well-known case of strong interaction amplitudes.  
Possibly the name of `quark-hadron duality' captures the essence of this `new' kind of duality. On the average, it is expected to work locally, i.e., for reasonable finite intervals of $s$, and in a sense this `quark-hadron duality' can be considered as a precursor of more accurate and important developments such as the SVZ-sum rules that appeared once QCD was proposed. 

\section{Quark-hadron duality, sum rules and the `new' Vector Mesons}

Up to now, our discussion of quark-hadron duality has been restricted to processes involving $u$, $d$ and $s$ quarks and the corresponding nonets of SU(3) vector mesons. New applications appeared as soon as new flavours were discovered and the value of $R$ was correspondingly increased. Just after the discovery of the $J/\psi$ resonances, M.~Greco with C.~A.~Dom{\'{\i}}nguez~\cite{ref:dg} --and some time later with J.~Pancheri and Y.~Srivastava~\cite{ref:gps}, taken now radiative corrections into account-- applied quark-hadron duality to estimate the 
averaged increase in $R$ produced by the new charmed quark. The results of these papers, 
\begin{eqnarray}
R &=& R_{u, d, s} + R_{\rm charm} \nonumber \\ &\simeq& \;\; 2.5 \; \; \,+ 1.2 {\rm \,\, (no\, rad. \, corrections)} \nonumber \\
&\simeq& \;\; 2.5 \; \;\,+ 1.8 {\rm \,\, (with \, rad.\, corrections)} ,
\end{eqnarray}
are in good agreement with the available data for $R$ in the $\sqrt{s}$ region between around 3 GeV (where the charmed channel opens) and 10 GeV (where the $b$-channel starts). 

In 1978, when the $\Upsilon$ resonances appeared above $\sqrt{s} \simeq 10$ GeV but the electric charge of the $b$-quark was not firmly established, 
M.~Greco~\cite{ref:mg2} used again duality ideas to estimate the further increase in $R$ in the new energy region predicting the decay width  
\begin{eqnarray}
\Gamma \left(\Upsilon (b \bar{b}) \to e^+ e^- \right) \simeq 1.2 \, {\rm keV},
\end{eqnarray}
which favours a $b$-quark electric charge of -1/3. 

Other, more formal developments were also considered. From canonical trace anomalies of the energy-momentum tensor, 
E.~Etim and M.~Greco~\cite{ref:eg2} derived the following general, $n$-valued family of quark-hadron duality sum rules
\begin{eqnarray}
\int_{s_0}^{\bar{s}} ds s^n { Im} \Pi (s) = {\alpha R \over 3}{\bar{}s^{n+1}\over n+1} - {c_n \over n+1},
\end{eqnarray}
where the limits of integration define the region where the two dual contributions are averaged and the imaginary part of the vacuum polarization function is related to the $e^+ e^-$ annihilation cross section into hadrons via 
\begin{eqnarray}
{ Im} \Pi (s) &=& {s \over 4 \pi \alpha} \sigma_{\rm had}(s) \\
&=& {4 \pi^2  \alpha} {m^2_\rho \over f^2_\rho} \sum_n \delta (s-m^2_n)
\end{eqnarray}
if one adopts the narrow width approximation for vector resonances in the final expression. 

In particular, for $n=0$ the preceding equation takes a much more familiar and simple  form,  
\begin{eqnarray}
\int_{s_0}^{\bar{s}} ds \left( { Im} \Pi (s) - {\alpha R \over 3} \right) =0, 
\end{eqnarray}
whose generalization to axial-vector channels and to channels with open strangeness was also discussed by  
E.~Etim, M.~Greco and Y.~Srivastava in~\cite{ref:egs2}. 

As previously stated, both the general form of these sum rules and their extension to different channels suggest that a part of the subsequent work leading to the extremely successful SVZ- or QCD-sum rules of 1979 was done by a reduced group of people under Mario's direction. 
Certainly, this initial ideas developed around 1972 were quite simple and naive --no gluons, no condensates could be invoked during those pre-QCD days-- but the central point, namely, the dual behaviour between quark- and resonance-contributions, was already there. 
For the present author it has been a great pleasure to remember those days as a postdoc member of the group enjoying a wonderful stay in Frascati. 


\acknowledgments
Thanks are due to Gino Isidori for his kind invitation to participate in the special Session Honoring Prof. Mario Greco and to the organizers of the meeting in La Thuile 2011 for the excellent atmosphere and organization.

\end{document}